\documentclass[11pt]{article}
\usepackage[latin1]{inputenc}
\usepackage[english]{babel}
\usepackage[namelimits]{amsmath}
\usepackage{amssymb}
\usepackage{amsmath}
\usepackage{amsthm}

\begin{document}
\title{First law of thermodynamics for dynamical apparent horizons and the entropy of Friedmann universes}
\author{
Stefano Viaggiu,\\
Dipartimento di Matematica,
Universit\`a di Roma ``Tor Vergata'',\\
Via della Ricerca Scientifica, 1, I-00133 Roma, Italy.\\
E-mail: {\tt viaggiu@axp.mat.uniroma2.it}}
\date{\today}\maketitle
\begin{abstract}
Recently, we have generalized the Bekenstein-Hawking entropy formula for black holes embedded 
in expanding Friedmann universes. In this letter, we begin the study of this new formula to obtain the first law of thermodynamics for 
dynamical apparent horizons. In this regard we obtain a generalized expression for the internal energy $U$ together with a distinction 
between the dynamical temperature $T_D$ of apparent horizons and the related one due to thermodynamics formulas. Remarkable,  when 
the expression for $U$ is applied to the apparent horizon of the universe, we found  that this internal energy is a 
constant of motion. 
Our calculations thus show  
that the total energy of our spatially flat universe including the gravitational contribution, 
when calculated at the apparent horizon, is an universal constant that can be set to
zero from simple dimensional considerations. 
This strongly support the holographic principle.
\end{abstract}
{\it Keywords}: Bekenstein-Hawking entropy; black holes; expanding universes; internal energy.\\
%PACS number(s): 04.70.Bw, 04.70.-s, 04.70.Dy, 04.20.Cv.

\section{Introduction}
Since of the Hawking's discovery that black holes radiate \cite{1}, his entropy formula represented a cornerstone for a 
sound formulation
of quantum gravity.
In an asymptotically flat spacetime, the black hole entropy $S_{BH}$ is proportional to its proper area
$A$, i.e.  $S_{BH}=\frac{k_B A}{4 L_P^2}$, where
$k_B$ is the Boltzmann constant and $L_P$ is
the Planck length. However, in a more realistic situation,
black holes are embedded in our expanding non asymptotically flat universe.
When one consider spacetimes that are not asymptotically flat, as for example a cosmological context,
the situation becomes rather more involved. First of all, in a cosmological context the identification of a black hole is not simple and 
only few exact solutions are known \cite{3,4,5,6,6a,T1,T2,T3,T4,T5,T6} as possible solutions describing black holes in expanding universes.
In this context, it has been shown \cite{7,8} that a fundamental ingredient is provided by the existence of the (outer) 
apparent horizon, rather than the teleological event horizon.
Despite the mathematical difficulty to obtain exact solutions, black holes 
certainly exist in our universe. As pointed in \cite{9}, to the best of my knowledge,
in practically 
all papers concerning black hole entropy in expanding universes, the formula $S_{BH}=\frac{k_B A}{4 L_P^2}$ is used 
as an ansatz.
 
A typical issue concerning black holes thermodynamics is the introduction of a volume term in the first
thermodynamic law. In the static case this term is not present. However, in an expanding universe, the possible existence of 
non static dynamic horizons reasonable requires the presence of a work term. 

In the usual approach, this 
work term is introduced by hand, in order to recast the Einstein's equations in a form similar to the first thermodynamic law
(see for example \cite{15,17,ac}.

In this paper we use a different approach.
In \cite{9} it is shown that the use of suitable theorems for the formation of 
black holes in expanding universes \cite{10,11,12,12c} leads to a more general expression for $S_{BH}$. In particular, 
an added term
proportional to $VH$ arises
that can be seen as due to the dynamical degrees of freedom of the expanding universe.
This paper is devoted to a study of some interesting physical consequences of the new expression for 
$S_{BH}$. In this new proposal, a volume (work) term naturally arises, without invoking analogies between Einstein's 
equations and thermodynamic laws. Moreover,
the new black hole entropy formula give us the possibility to write a generalised expression for the first law of thermodynamics in a simple 
and physically clear form.\\ 
Finally, as it is customary in the literature, we can apply our generalized entropy formula to the apparent horizon 
(Hubble radius for spatially flat Friedmann solutions) of the universe. This permit us to investigate, in a simple manner, the 
thermodynamic properties of our universe.

In section 2 we present the new formula together with the preliminaries set up of this work. In section 3 we write down our generalized first
law of thermodynamics.
Section 4 is
devoted to the application of the new entropy formula to the whole universe. In section 5 we calculate the entropy of
closed and hyperbolic Friedmann universes. Finally, section 6 collects some final remarks and conclusions.

\section{Preliminaries}
In \cite{9} has been shown that, after taking into account the suitable theorems in \cite{10,11,12} 
for the formation of trapped surfaces
in a cosmological context, a modification of the usual formula for the Bekenstein-Hawking entropy does appear. 
The argument
used is similar to the one present in \cite{2} and \cite{13} (entropy spherical bound, see also \cite{9} for a proposal of generalization).
The Bekenstein bound states that there exists an universal bound for the
entropy $S$ of a spherical object of radius $R$ and energy $E$ given by 
$S\leq S_{max}=\frac{2\pi k_B RE}{\hbar c}$. The entropy bound \cite{13} is weaker than the Bekenstein one and it is a 
consequence of the requirement that a system be gravitationally  stable, otherwise a black hole forms 
after a gravitational collapse.
As well known, 
the identification of the event horizon is a global property of the spacetime and its identification is a formidable task. A 
more manageable (local) ingredient is provided by trapped surfaces. 
A more simple approach is to study the conditions for the formation or non formation of trapped surfaces. 
In fact in \cite{isr} it has been shown that the presence of trapped surfaces 
caused by spherically symmteric mass-energy concentration satisfying the weak energy condition, unavoidably leads to a 
black hole, at least in the static case.

To start with, we quote the theorems present in \cite{10,11,12,12c} 
(in this and the following two sections we quote only the spatially flat case). We thus restrict our attention to the spherical case.
We denote with $\Sigma$ the spherical three-dimensional hypersurface obtained at $t=const$.
The initial data set $g_{ab}, K_{ab}, \rho$ are given on the spatial hypersurface $\Sigma$, where 
$g_{ab}$ is the metric, $K_{ab}$ the extrinsic curvature and $\rho$ the energy-density of the spherically perturbed 
Friedmann flat spacetime on the three-dimensional space $\Sigma$.
We  also denote with $J_a$ the matter current and with $\rho={\tilde{\rho}}+\delta\rho$, where 
${\tilde{\rho}}$ is the unperturbed density and $\delta\rho$ the spherically symmetric energy-density perturbation.
A trapped surface $S$ is a compact two-dimensional space-like surface with an outgoing flux $\theta$ of future-directed null
geodesics orthogonal to $S$ and everywhere negative.
Suppose that $J_{a}=0$ (no matter current flowing
through $\Sigma$) and that $K_{a}^{a}=const$, i.e. the rate of change of the volume of $\Sigma$ is not perturbed: if for a sphere $S$
with proper mass excess $\delta M$ we have
\begin{equation}
\delta M\frac{G}{c^2}<\frac{L}{2}+\frac{A H}{4\pi c}, 
\label{1z}
\end{equation} 
then $S$ is not trapped. 
As a result, the theorem assures the non existence 
of trapped surfaces for spacetimes with a mass concentration on spacelike surfaces 
with a pure trace extrinsic curvature.  This is not a restrictive hypothesis.
For example for the McVittie solution \cite{3} the extrinsic curvature is $3H(t)$ on comoving
foliation. It is important to note that in order to obtain the inequality (\ref{1z}), no linear approximation is invoked
and backreaction is taken into account.
In practice, 
thanks to the spherical symmetry of the mass-excess $\delta M$, the spatial part of the three metric describing the spacetime
with the pertubation can be exactly represented as a conformally flat metric in isotropic coordinates. This allows exact computations.

Inequality (\ref{1z}) is the starting point of our considerations. First of all,  
by using the Bekenstein bound in the spherical case $S\leq S_{max}=\frac{2\pi k_B RE}{\hbar c}$ 
(see \cite{9}) with the inequality (\ref{1z}) with $H=0$ we obtain the usual entropy formula for static asymptotically flat spacetimes. In this regard, the proper length $L$ is identified with the proper length of event horizon of the black hole.\\
The same reasoning can be done with $H\neq 0$. In practice, we use the same Bekenstein bound 
$S\leq S_{max}$ of the stationary 
asymptotically flat case, but with the bound on the mass-energy given by (\ref{1z}). To this purpose
in an expanding universe, as well known,
it is not a simple task to identify the event horizon of a black hole. As suggested in \cite{7,8} (see also \cite{ZL}), 
black holes in expanding universes are defined  by the existence of apparent horizons. Hence, we can identify $L$ with the
outer apparent horizon $L_h$ of the black hole. We obtain: 
\begin{equation} 
S_{BH}=\frac{k_B A_h}{4 L_P^2}+\frac{3k_B}{2c L_P^2}V_h H,
\label{1}
\end{equation}
where  $A_h$ and $V_h$ denote the proper area and the (effective geometrical) volume 
of the black hole at apparent horizon. 
Moreover,  
according with the prescription made in \cite{14} for the effective volume of a spherical black hole,  
by denoting with $L_h$ the proper 
length of the apparent horizon we have  $L_h=\sqrt{\frac{A_h}{4\pi}}$ and 
$V_h=4\pi L_h^3/3$. The meaning of 
(\ref{1}) is simple. The entropy is a measure of the energy content in a given region.
Since, as evident from the formula (\ref{1z}), the expansion of the universe makes more difficult to build a black hole, more energy (entropy)
with respect to the static case ($H=0$)  can be enclosed in a sphere of proper radius  $L$. 

Physically, since the entropy
of an object can be seen
as a measure of its degrees of freedom, the term proportional to $VH$ in (\ref{1}) is a 
consequence of the degrees of freedom related to the non static nature of the spacetime. This conclusion is enforced by the relation among this term and the
Cardy-Verlinde entropy formula. As shown in \cite{9}, by considering the Bekenstein bound with the correct normalization 
for closed Friedmann universes, our added term is exactly equivalent to the Cardy-Verlinde entropy. In fact, in the 
usual thermodynamics, the vibrational and rotational motions of the molecules composing a gas or a solid enter in the
computation of the degrees of freedom. Similarly, the non static nature of an expanding universe caused by the
Hubble flow enters in the determination of the degrees of freedom of a black hole. 

Remember that the formula (\ref{1}) has been obtained in a spherical context, where $A_h$ and 
$V_h$ are not obviously  independent quantities. However, we can use the theorem present in \cite{11} to extend the 
expression (\ref{1}) to non-spherical black holes. Only in this case $A_h$ and $V_h$ are independent variables. The expression
(\ref{1}) can be also generalized to black holes embedded in Friedmann cosmologies with positive (closed) and 
negative (open) curvature. According to all cosmological data, we live in a spatially flat universe
and thus we mainly analyze this case. The generalization to the other Friedmann cases will be briefly discussed 
in section 5.

\section{First law of thermodynamics for dynamical apparent horizons}

In the ordinary thermodynamics, it is well known that the Euler's homogeneous theorem functions allows to write
the relation $U=TS-PV$ (we consider for simplicity the case with constant number of particles $N$). 
The Euler's relation above is satisfied  
provided that $U,S,V$ are extensive quantities. The extensivity of the internal energy $U$ in ordinary thermodynamical systems is a consequence of the short range interaction between particles. The situation changes drastically when gravity
(long range interaction) comes into action. In fact, in a general relativistic cosmological context we expect an interaction term
$U_{int}$ (similarly to the Newtonian gravitational potential)
that cannot be neglected. The internal energy cannot be considered in a general relativistic framework
as an extensive quantity and  thus the 
Euler's relation $U=TS-PV$ breaks down. We must use a different strategy to obtain the differential form of the first law from a finite statement. 
To start with, after taking the total differential of (\ref{1}) we obtain:
\begin{equation}  
dS_{BH}=\frac{k_B}{4L_P^2} dA_h+\frac{3k_B}{2c L_P^2} V_h dH+
\frac{3k_B}{2c L_P^2}H dV_h.
\label{2}
\end{equation}
Note that in the equation (\ref{2}) a work-like term naturally originates due to the added one in (\ref{1}).
Usually, the entropy $S$ of a system is a function of the internal energy $U$ and of the volume $V$ and thus:
\begin{equation} 
dS(U,V)={\left(\frac{\partial S}{\partial U}\right)}_V d U+{\left(\frac{\partial S}{\partial V}\right)}_U d V=
\frac{d U}{T}+\frac{P}{T} d V.
\label{3}
\end{equation} 
To specify the thermodynamic parameters $U$ and $P$, 
we must obtain the expression for $T$. 
This is not a trivial task. In fact, in a non stationary cosmological context the timelike coordinate it is not a Killing isometry. Moreover,
in a cosmological context event horizons become teleological objects that are attainable only for eternal observer.
Fortunately, in \cite{7,8} it has been shown that  the relevant concept for identifying black holes in non-stationary contexts is provided by 
the apparent horizon together with its surface gravity parameter $k_h$
(see also \cite{T6,H3}), defined in terms of the Kodama vector field.
Consider (see \cite{H3}) a general spherically
symmetric metric in the form:
\begin{equation}
ds^2={\gamma}_{ij} dx^i dx^j+R^2(x^i)d{\Omega}^2,\;\,\;i,j\in\{0,1\},
\label{H1}
\end{equation}
where $R$ denotes the proper radial radius.
The location of the apparent horizon $L_h$ is given by the equation ${\gamma}^{ij}R_{,i}R_{,,j}=0$, where commas denotes partial
derivative. The dynamical surface gravity $k_h$ and the related dynamical temperature $T_{D_h}$
calculated at $L_h$ are thus given by:
\begin{equation}
k_h={\left[\frac{1}{2\sqrt{-\gamma}}{\left(\sqrt{-\gamma}{\gamma}^{ij}R_{,j}\right)}_{,i}\right]}_{|L_h},\;\;
T_{D_h}=\frac{c\hbar|k_h|}{2\pi k_B}.
\label{H2}
\end{equation} 
A useful formula has been derived in \cite{H3}. By denoting with $T^{(2)}$ the trace of the energy-stress tensor on the 
normal space $\{0,1\}$ calculated at $R=L_h$, one obtain for $k_h$ on the dynamical horizon the expression:
\begin{equation}
k_h=\frac{1}{2 L_h}+4\pi\frac{G}{c^4} L_h T^{(2)},
\label{H3}
\end{equation}
It should be noticed that the thermodynamic temperature is defined as
${\left(\frac{\partial S}{\partial U}\right)}_{V_h}=1/T$. The partial derivative with respect to the internal energy must be calculated
keeping the proper volume $V_h$ (i.e. $L_h$) of the dynamical apparent horizon fixed. We can define a more appropriate notion 
of temperature when usual thermodynamc relations are used. To this purpose, we denote with $T_{L_h}$ the 
'thermodynamic' temperature defined  as $T_{L_h}=T_{D_h}(L_h=const)$. As a result, we can still use expression
(\ref{H3}) for $k_h$ but with $T^{(2)}$ calculated at $R=L_h=const$, 
$T^{(2)}(L_h=const)=T_{L_h}^{(2)}$. 
Moreover, we can impose the usual Clausius relation
$T_{L_h}dS_{BH}=\delta Q_h$. In a similar way of \cite{H3},
we use the practical rule to identify the terms $\sim L_h^2 dL_h$ as terms proportional to
$dV_h$. For the first law we obtain:
\begin{eqnarray} 
& &\delta Q_h=dU_h+P_h dV_h,\label{H4}\\
& & dU_h=\frac{c^4}{2G} dL_h+dH
\left[\frac{c^3}{2G}L_h^2+\frac{4\pi}{c}L_h^4 T_{L_h}^{(2)}\right],\label{H5}\\
& & P_h=T_{L_h}^{(2)}\left[1+\frac{3H}{c}L_h \right]+\frac{3c^3 H}{8\pi G L_h} \label{H6},
\end{eqnarray}
where with $U_h$ and $P_h$ we denoted respectively the internal energy and the 
'effective' pressure calculated at $R=L_h$. In (\ref{H6}), the term $P_h dV_h$ is a work term $dW_h$. This work term 
$dW_h$ has been written in the usual way as $P_hdV_h$, although generally the 'effective pressure' has a more general expression than the 
usual one $P$ present in the energy-momentum tensor (and present in $T_{L_h}^{(2)}$). 
As pointed in \cite{H4},  the fact that the pressure of the work term is not merely the one present in the stress tensor, can be related 
to the general non comoving nature of apparent horizons.\\
Note that the expression (\ref{H5}) can be used to calculate the variation of $U_h$ after a given thermodynamical 
transformation. To calculate the variation of $U_h$ caused by the expansion of the universe, we must to integrate 
the (\ref{H5}) between an initial time $t_i$ and a final time  $t_f$.\\
By a first inspection of the internal energy (\ref{H5}), we recognize with the term $c^4 L_h/(2G)$ nothing else but the Misner-Sharp energy
evaluated at the apparent horizon. In our frame, a further term does appear proportional to $dH$. 
First of all, as shown in the appendix 1,
our tractation can include the de Sitter cosmological spacetime (without including its static patch, such as the 
Schwarzschild de Sitter black hole solution). 
The term $\sim dH$ in  (\ref{H5}) is vanishing
for $H=const$, i.e. in the de Sitter (non static) cosmological spacetime where $H=c\sqrt{\Lambda/3}$, 
and in the stationary case  ($H=0$). This further term can be associated to the non local gravitational energy due to the 
non static nature of the universe, and more precisely to the non static nature of the apparent horizon in an expanding universe.
This term is absent in the usual expression for the internal energy (including only the Misner-Sharp term) and, as shown in the next section,
will play an important role assuring the conservation of the total energy at the apparent horizon of the universe. The 
proposed first law given by
(\ref{H4})-(\ref{H6}) generilizes the one present in \cite{H3} concerning black holes in expanding universes, thanks to the added term $\sim VH$
of our proposal.

In the literature, it is rather usual to use for black hole entropy also in expanding universes (see for example \cite{15} and
references therein) the expression $S\sim A/4$. 
Our proposal is a consequence of theorems of general relativity suitable for the black hole formation in expanding 
Friedmann universes. 

To study the thermodynamics of the universe it is customary to consider
the apparent horizon $L_h$ equipped with temperature 
$T\sim 1/(2\pi L_h)$ and entropy $S\sim A/4$, mimicking the thermodynamics of stationary black holes. In any case, cosmological 
expanding universes have apparent horizons and are a natural arena to apply the formalism of this paper with the related surface gravity $k_h$ given by (\ref{H2}). Another issue when one consider the cosmological case is related to the bound of the entropy.
In a cosmological context the bound becomes dynamical (see \cite{d5,bb2,bb3,bb4}). As an example, 
in the proposal present in 
\cite{d5,bb2} it is shown that, in order to save the holographic principle with the 
expression $S\sim A/4$, the bounding area is not the one enclosing a certain region 
$A$ but rather the one given by the light-sheet of $A$ itself. It is well known that the usual entropy bound 
fails in a cosmological context. In \cite{9} it has been shown that the added term in the entropy proportional
to $VH$ can saturate the entropy bound. In any case, both the entropy (\ref{1}) together with the usual expression 
$S\sim A/4$ used in the literature have the status of proposals.\\ 
Note that the holographic principle \cite{13} is based 
on the idea that the maximum entropy of a certain volume $V$ is provided by the largest black hole fitting inside $V$.
As shown by the theorem (\ref{1z}), the expansion of the universe makes more difficult to build a black hole and as a consequence 
more energy and entropy can be enclosed  within $V$ with respect to the static case. 
If this were not the case, we could have an entropy $S_h\sim A_h/4$ for the apparent horizon $L_h$
of our universe but at the same time we could have a spherical object of radius $L_h$ with an entropy $S_m$ greater
than $A_h/4$ but less than $S_{BH}$ (equation (\ref{1})) without forming a black hole, i.e. an evident contradiction with the
holographic principle.

\section{An application: first thermodynamics law at the Hubble radius}

In the usual view, 
Friedmann equations together with the first law of thermodynamics applied to comoving spheres of volume $V_c$ imply that 
$TdS=\delta Q=0$ $dU=-P dV_c$ (see for example \cite{H4}), where $P$ is the hydrostatic pressure present in $T_{\mu\nu}$.
However, as pointed in \cite{20},
the universe contains a large amount of entropy and this fact seems to be in contradiction with an adiabatic 
isoentropic universe.\\
A more suitable locus to test thermodynamics laws has been shown to be the apparent horizon \cite{7,8,15,17,ac,20,18}.
In this regard, as stated at the end of the section 3, the apparent horizon is the natural locus where we can apply the 
machinery of this paper.
As an example, it is customary (see \cite{e1,e2,e3} in the context of the so called 
'entropic cosmology'  \cite{15,17,20})
to associate at the apparent horizon of the universe an entropy and a temperature. 
Tipically, for the entropy
the expression $S\sim A_h/4$ evaluated at the apparent horizon (in a flat space)
$R_h=c/H$ of the universe is used. This is justified from the fact that at the particle horizon or the future event horizon 
the thermodynamics is ill-defined (see for example \cite{H4} and references therein). 

In \cite{15} the authors show that by assuming that the apparent horizon of the universe is equipped with a temperature $T\sim H$ and an entropy $S\sim A/4$, then the 
differential form of the Friedmann equations is equivalent to the first law with $\delta Q=TdS=dU$. Moreover, in \cite{ac} the authors
write down the differential form of the Friedmann equations as a first law in the following way:
\begin{equation}
T_{D_h}dS=dU_h+W_h dV_h,\;\;\;W_h=\frac{P-\rho c^2}{2},
\label{HH6}
\end{equation}
where $W_h$ is  the coefficient of the work term and $U_h=c^4L_h/(2G)$ is the Misner-Sharp energy term and 
\begin{equation}
T_{D_h}=\frac{\hbar c}{2\pi k_B L_h}\left |1-\frac{L_{h,t}}{2H L_h}\right |
\label{H7}
\end{equation}
is the temperature at the apparent horizon and show that this is consitent with an entropy $S\sim A/4$.
In deriving the expression (\ref{HH6}) it is assumed for the temperature the expression
(\ref{H7}) with the term $L_{h,t}$. However, as stated in the section above, the thermodynamic temperature is 
given by ${\left(\frac{\partial S}{\partial U}\right)}_{V_h}=1/T$ calculated at
$V_h=const$, but for the apparent horizon this means $L_h=const$ and consitency with the differential form of the first law
would require to take for the temperature  the expression
\begin{equation} 
T_{L_h}=\frac{\hbar c}{2\pi k_B L_h}.
\label{H8}
\end{equation}  
Espression (\ref{H8}) is the one often used as the temperature of the apparent horizon, but in our context it is the consequence of the ordinary formulas of thermodynamics and not the outcome of an approximation.\\
Moreover, it is rather questionable that the Misner-Sharp mass is a good expression for the internal energy $U$
in an expanding non-static
non asymptotically flat spacetime. In fact, the Misner-Sharp mass at the apparent horizon $L_h$
is nothing else but $\rho V_h$, where $\rho$ is the matter density
and $V_h$ the volume of the apparent horizon, i.e.
the Misner-Sharp mass represents the matter content inside $V_h$. The 
contribution given by the gravitational energy is non-local and cannot be included in the energy-momentum tensor $T_{\mu\nu}$. 
It is reasonable that to an expanding universe can be associated a gravitational energy due to the expansion that should be included in the expression for $U$. Finally, there is no reason to believe that Einstein's equations are thermodynamical equations, in the same way as 
classical equations of motion do not contain thermodynamics laws. Thermodynamics captures properties of the matter that are not 
described by ordinary mechanical laws. 

For the reasons above we use a different approach following the procedure of section 3.
As a consequence, for the expression 
(\ref{H5}),  (\ref{H6}) and thanks to (\ref{H8}) we have:
\begin{equation}  
\Delta U_h=\frac{c^4}{G}\Delta L_h+\frac{c^3}{G}\int_{t_1}^{t_2} L_h^2 H_{,t}dt,\;\;\;P_h=\frac{3 c^3 H}{4\pi G L_h}.
\label{H9}
\end{equation} 
The pressure $P_h$ in (\ref{H9}) is not the hydrostatic pressure $P$ 
present in the stress energy tensor $T_{\mu\nu}$: it is the pressure that originates the 
work term due for a dynamical apparent horizon. Moreover, it should be noted that 
$P_h/T_{L_h}={\left(\frac{\partial S}{\partial V}\right)}_{U_h}$: 
this partial derivative must be calculated at the apparent horizon 
$R=L_h$ for $V_h\neq const$ and for $U_h=const$. 
When we evaluate (\ref{H4})-(\ref{H6}) at the apparent horizon (Hubble radius)
$L_h=c/H$, we have $dU_h=0$, i.e. $U_h$ is a constant of motion at the apparent horizon. 
In practice, the form of the entropy
given by (\ref{1}) together with our definition for the 'thermodynamic' temperature $T_{L_h}$ conspire to obtain the desired result.
We thus obtain:
\begin{equation} 
dQ_h=T_{L_h} dS_h=P_h dV_h,\;\;T_{L_h}=\frac{\hbar H}{2\pi k_B},\;\;P_h=\frac{3c^2 H^2}{4\pi G}.
\label{H16}
\end{equation} 
with respect to the equation (\ref{H16}) the universe cannot be considered isoentropic and 
$dU_h=0$, as happens for an ideal gas
in a free expansion.
The expression (\ref{H16}) for $dQ_h$ can be written as:
\begin{equation} 
dQ_h=-\frac{3\hbar c^5}{2\pi k_B G T_{L_h}^2}dT_{L_h}.
\label{H19}
\end{equation}
Since the internal energy $U_h$ is a conserved quantity at the Hubble radius, we can define the specific heat 
at constant $U_h$:
\begin{equation}
C_{U_h}={\left(\frac{dQ}{dT_{L_h}}\right)}_{U_h}=-\frac{3\hbar c^5}{2\pi k_B G T_{L_h}^2}.
\label{H20}
\end{equation} 
The specific heat $C_{U_h}$ is negative. This happens because, when the universe expands, $dT_{L_h}<0$ and hence
$dQ_h>0$, i.e. the temperature decreases but the entropy of the universe increases. This can be seen by integrating
the (\ref{H19}) between the initial temperature $T_{L_{h1}}$ up to the final one $T_{L_{h2}}$:
\begin{equation}
Q_h(T_{L_{h2}})-Q_h(T_{L_{h1}})=\frac{3\hbar c^5}{2\pi G k_B}\left(\frac{1}{T_{L_{h2}}}-\frac{1}{T_{L_{h1}}}\right).
\label{H21}
\end{equation} 
Consider a universe filled
with a matter-energy content satisfying the weak and the strong energy conditions. In this case, 
during the cosmological expansion $T_{L_{h2}}-T_{L_{h1}}<0$ and hence 
$Q_h(T_{L_{h2}})-Q_h(T_{L_{h1}})>0$: the universe during the expansion gains heat, i.e. entropy increases. This means that the universe absorbs energy (heat) and this energy is flowing inwards from the Hubble horizon. The idea that the universe generates entropy can be traced back to \cite{19}. More recently, in 
\cite{20} has been shown that a Hawking-like radiation at the Hubble radius can be a viable mechanism for inflation.
This radiation, differently from the Hawking one that is escaping to infinity, is ingoing, as happens thanks to
formula (\ref{H21}). This reasoning shows that there exists some analogy with the mechanism depicted in \cite{20}. 
Moreover, note that for an inflatting universe violating the strong and the dominant energy conditions 
($H_{,t}>0$), from (\ref{H21}) we have
$\Delta Q_h<0$, and hence the entropy decreases, while a de Sitter universe evolves adiabatically with 
$S_h=const$.

As a final consideration, we analyze the hypothesis that the cosmological dynamical term in
(\ref{1}) is due to the non local contribution of the gravitational expansion energy. To this purpose,   
it is well known 
that the role of the gravitational energy at large cosmological scales is
not yet well understood.\\ 
The first term in (\ref{H5}) is nothing else but the Misner-Sharp energy $E_{ms}=c^4L_h/(2G)$. 
When one considers this term
at the apparent horizon of a Friedmann solution, we have $E_{ms}(L_h=c/H)=\rho(t)c^2 V_h,\;V_h=4\pi c^3/(3H^3)$. 
Obviously, thanks to the expansion of the universe, this quantity increases with time ($E_{ms}\sim 1/H$). 
Since we found a conservation law for the energy at the Hubble radius of a pure
Friedmann flat universe, the terms proportional to $dH$ in (\ref{H5}) must be considered as the non local contribution $E_{grav}$
(not included in $T_{\mu\nu}$) to the internal energy due to the gravitational expansion energy. For ordinary cosmologies with matter field
satisfying the energy conditions, this term is decreasing in time and at the apparent horizon $L_h=c/H$ we have
${\left(E_{ms}+E_{grav}\right)}_{|L_h}=K=const$. 
The fact that the conservation law arises only at the Hubble radius of a Friedmann universe
further support the idea that apparent horizons are privileged locus for the thermodynamics and also the 
validity of the holographic principle.
Also note that the same happens for a de Sitter cosmological spacetime, where $dH=0$ and the Misner-Sharp
energy term is constant since $E_{ms}\sim L_h\sim 1/\sqrt{\Lambda}$.
The value of $K$ is an important open question
regarding the origin of our universe. A zero value for $K$ would imply a universe born from a vacuum state of zero energy.
To this purpose, note that in a de Sitter cosmological spacetime we have $K=c^4/(2G)\sqrt{3/\Lambda}+E_v$,
where $E_v$ is a new constant that can depend on $\Lambda$, 
i.e. $E_v(\Lambda)$.
Since our universe becomes for $t\rightarrow\infty$ asymptotically de Sitter ($dH\rightarrow  0$ in (\ref{H5})), 
we have $K=c^4/(2G)\sqrt{3/\Lambda}+E_v$. Moreover, the Schwarzschild spacetime has total 
energy given by $mc^2$ ($m$ is the well known ADM mass of the black hole), we expect the Minkowski space to have zero energy,
since for $m\rightarrow 0$ we obtain the Minkowski metric.
Moreover, Minkowskian spacetime can be obtained from the de Sitter one in the limit $\Lambda\rightarrow 0$. As a consequence
we must have: $E_v=-c^4/(2G)\sqrt{3/\Lambda}+C(\Lambda)$, where $C(\Lambda=0)=0$. 
We can fix the value of the constant
$C$ from simple dimensional considerations.
If $C$ is independent from $\Lambda$, then $C=0$, i.e. our universe has the same energy (at the apparent horizon)
of the Minkowski space and of the de Sitter cosmological metric. 
Otherwise, $C$ can be only a function of $\Lambda$ and of the fundamental constants $G$ and $c$, i.e. 
$C=C(\Lambda,G,c)$. From simple dimensional arguments it is easy to see that $C\sim 1/\sqrt{\Lambda}$. Hence, 
the only possibility to obtain $C(\Lambda=0)=0$ is that $C=0$. 
These reasonings do not hold only by introducing the 'extra' quantum 
constant $\hbar$ and in this case $C\sim \hbar c\sqrt{\Lambda}$
(at the appendix 2 we give some arguments relating the presence of the non-classical 
constant $\hbar$ to the dark energy).\\
As a consequence, at the apparent horizon, 
according to an old conjecture
firstly proposed in \cite{e0} and reproposed later (see \cite{F1} and reference therein),
a spatially flat 
Friedmann universe has the same zero energy of the Minkowski spacetime. According to this conjecture, only in the spatially flat 
case we obtain this result. It is rather reassuring that exactly at the apparent horizon, where a dynamical surface gravity $k_h$
can be defined, for a Friedmann 
flat universe we obtain a vanishing internal energy, as expected. This fact strongly supports our proposal
(\ref{1}).

\section{Entropy for closed and hyperbolic Friedmann universes}

In this section, we calculate the entropy for hyperbolic and closed Friedmann universes. To this purpose, we quote the theorems present in
\cite{10,11,12c} where conditions for the formation of trapped surfaces caused by spherical perturbations in 
hyperbolic and closed Friedmann universes are given. 
In this regard, we can use the same arguments of section 2 to obtain the expression (\ref{1}). Under the same conditions of the theorem 
(\ref{1z}), if for a spherical surface $S$ with area $A$ and volume $V$ and with  proper mass excess $\delta M$ we have:
\begin{equation}
\frac{G}{c^2}\delta M < \frac{L}{2}+\frac{AH}{4\pi c}-k\frac{3V}{8\pi a(t)^2},
\label{all}
\end{equation} 
then $S$ is not trapped. For $k=0, -1, +1$ we have respectively the flat, the hyperbolic and the closed Friedmann background.
With $a(t)$ we denote the scale factor of the spatial slice $\Sigma$.\\ 
By applying again the entropy bound for a spherical mass concentration we obtain:
\begin{equation}
S_{BH}=\frac{k_B A_h}{4 L_P^2}+\frac{3k_B}{2c L_P^2}V_h H-
\frac{3k k_B}{4L_P^2}\frac{L_h V_h}{a(t)^2},
\label{alla}
\end{equation}
Note that, according to physical intuition, for an open universe the further term with respect to the flat case (\ref{1}) is positive
(in an hyperbolic universe it is more difficult to build black holes with respect to the flat $k=0$ case) and more entropy can be enclosed
in a given volume $V_h$. Conversely, in a closed universe it is more easy to build black holes and as a result the maximal entropy available in a same proper volume $V_h$ is less than the flat case. In the following, we apply the formula (\ref{alla}) to the apparent horizons of Friedmann 
universes. The proper radius of apparent horizons in Friedmann universes is given by:
\begin{equation} 
L_h=\frac{c}{\sqrt{H^2+\frac{k c^2}{a(t)^2}}}.
\label{al1}
\end{equation} 
As well known, the expressions for the proper volume $V_h$ for hyperbolic and closed Friedmann cosmologies are obtained by integrating
the volume element $dV=d\phi d\theta dr \sqrt{g^{(3)}}$ where as usual $g^{(3)}=a^6 r^4 {\sin}^2\theta/(1-kr^2)$. However, as stated in
\cite{H4}, the Misner-Sharp mass  within $V_h$ is given by $\rho V_h$, where $\rho$ is the matter density and
$V_h=4\pi L_h^3/3$ (areal volume), with $L_h$ given by (\ref{al1}). Hence, also for  closed and hyperbolic Friedmann universes the 'thermodynamical'
or geometric volume has the same expression of the flat case, but calculated at $L_h$ given by (\ref{al1}).\\
We can use the same strategy of the section above to obtain the differential form of the first law by performing the total differential
of (\ref{alla}). The expression of the thermodynamical temperature is again given by (\ref{H8}). 
In particulr, for the internal energy $U$
we obtain:
\begin{equation}
dU_h=\frac{c^4}{G}d L_h+\frac{c^3}{G}L_h^2 dH+
\frac{3kc^4}{4\pi G}\frac{V_h}{a(t)^3} da.
\label{al2}
\end{equation} 
Differently from the flat case, $dU_h$ given by (\ref{al2}) is not vanishing at $L_h$ given by (\ref{al1}): this happens only for $k=0$.
We found that the conservation of the internal energy $U$ at the apparent horizon is fulfilled only in the flat case.
This is due to the role of the curvature terms in (\ref{al2}) depending on $k$. To a negative curvature can be associated a positive 
'binding' energy (due to the term $-k c^2/a(t)^2$ in the Friedmann equations). This implies that more energy and thus more entropy can be 
enclosed within $V_h$.
Conversely, for a closed universe (strong attractive regime), this 'binding' term is negative and thus less energy (and entropy) 
than the other cases can be localized within  $V_h$. Since curvature is not present in Newtonian gravity, 
the contribution due to $k$ has no Newtonian counterpart and thus it is not unexpected that energy conservation holds at the apparent
horizon only in the flat case. For an expanding universe with matter content satisfying the weak energy condition, the apparent horizon is
increasing with time. This means that 'new space' is created: for $k=-1$  this implies the 'creation' of positive energy while for 
$k=1$ we have creation of negative energy. 
A zero curvature is associated to the
flat case and thus a zero binding curvature energy.

\section{Conclusions}
In this paper, we have investigated some interesting thermodynamics relations due to the  new proposal \cite{9}
for a generalized 
Bekenstein-Hawking entropy suitable for expanding universes, with particular evidence to the spatially flat,
the universes where we probably live (at least in in a statistical sense).
In this regard, we use a close analogy with ordinary thermodynamics. These analogies allow 
us to write down a new expression for the first thermodynamics law.

A first fundamental ingredient is a distinction between the dynamical temperature $T_{D_h}$ arising from the definition 
of the surface gravity for dynamical apparent horizons and the one $T_{L_h}$ more intrinsically related to the usual 
thermodynamics. In practice $T_{L_h}$ is obtained by calculating $T_{D_h}$ at constant $L_h$ 
(constant proper volume $V_h$ in the spherical case).
As a consequence, a suitable expression arises   
for the black hole internal energy $U_h$.\\
A very interesting property of this internal energy obtained with respect to $T_{L_h}$
is that, when applied to the 
whole universe, i.e. the Hubble radius (apparent horizon in the flat case), it is a constant of motion for the universe. 
A similar distinction can be found
in \cite{buc}.
Obviously, the dynamical temperature
$T_{D_h}$ is not directly related to our usual notion of thermodynamic temperature because a statistical definition for 
$T_{D_h}$ is still lacking. Nevertheless, the introduction of $T_{L_h}$ is required by a closest analogy with usual 
thermodynamics, in particular with the definition of the thermodynamical temperature requiring a partial derivative to be calculated
at constant volume $V_h$. As pointed in the paper, the fact that $U_h$ is a constant of motion at $R=L_h$ is an intriguing
properties with interesting physical consequences.
It is the conspiracy between our proposed entropy formula (\ref{1}) and
our definition of thermodynamic temperature $T_{L_h}$ (instead of the dynamical one $T_{D_h}$) that permits this interesting and reasonable 
result. This point is often missing in the literature. There, an important open issue is related to the meaning of the conservation law for 
$U_h$ at the apparent horizon. According to the interpretation given in \cite{9}  concerning
the term $\sim VH$ in (\ref{1}), we argue that 
the heat flow crossing inward the Hubble radius can be reasonable due to the non local expansion gravitational energy. This 
interpretation is enforced by the expression of the internal energy, thanks to the term proportional to $dH$ 
present in the  expressions (\ref{H5}) and (\ref{H9}) for $U_h$. In fact, since the first term in (\ref{H5}) for $U_h$ is the Misner-Sharp
energy, the added term could be as well 'pure' (expansion) gravitational energy. 
Regarding the suggestions present in the literature (see \cite{F1} and references therein)
that the total energy of the universe including the gravitational contribution could be zero,  
our calculations show that this is certainly the case for spatially flat Friedmann universes, but only at the apparent horizon of the universe. This further supports the idea that the apparent horizon is the right place to study the thermodynamics of the whole universe, i.e. the 
universe is thermodynamically equivalent to a system with internal energy $U$ ($U=0$ for the flat case) enclosed in a sphere of radius given by the apparent horizon.

The vanishing of $U$ at the apparent horizon
does not happen when one consider hyperbolic and closed Friedmann universes at the apparent horizon: in that cases, the non
vanishing curvature due to a non vanishing $k$ comes into action. This curvature contribution to $S_{BH}$ and $U$ has no
Newtonian counterpart. In practice, the non conservation of the total energy at the apparent horizon is a consequence of the
'creation' of new space due to the expansion of the universe. 

In any case, all these reasonings firstly show that the apparent horizon
can be considered as the proper radius of the universe, at least from the thermodynamics point of view. 
Moreover, from the results of this paper, it is clear that for a consistent formulation of the first law and thanks to the non-static nature of
the Friedmann universes, the internal energy $U$ cannot contain only the contribution from the matter (Misner-Sharp term)
but also the contributions due
to the gravitational energy caused by the expansion and the 'binding energy' due to a non-vanishing curvature (creation of space with time).
Only in the flat case $k=0$ (zero curvature energy) we have a perfect balance between the positive energy of matter, i.e.
the Misner-Sharp energy, and the gravitational expansion energy, according to the results of \cite{F1}. 
This can have interesting cosmological consequences. In fact, as firstly suggested in \cite{e0}, a universe with zero total energy can 
be emerged from quantum fluctuations of a Minkowskian spacetime. From this point of view, it is not a surprise that we live in a 
spatially flat universe (at least in a statistical sense), since a universe born from a Minkowskian spacetime must have zero total energy.       

\section*{Acknowledgements} 
I would like to thank Alessandra D'Angelo for stimulating discussions and for her constant encouragement, 
Luca Tomassini for interesting and useful discussions.

\section*{Appendix 1}
In this appendix, we extend the theorem in \cite{12} for the the de Sitter cosmological solution. The results below do not
include black holes in the static patch of the expanding de Sitter universe. As a consequence, the 
static Schwarzschild de Sitter solution is excluded from this tractation.

The inequality (\ref{1z}) is obtained in \cite{12} starting from an initial data set 
for the Einstein equations given by the four objects: the energy density $\rho$,  the unperturbed metric ${\tilde{g}}_{ab}$,
the extrinsic curvature ${\tilde{K}}_{ab}$ and the matter current density ${\tilde{J}}_a$ defined on a three-dimensional
spherically symmetric manifold $\Sigma$. These quantities cannot be arbitrary, but they must satisfy
the Hamiltonian and the momentum constraints. The momentum constraint is automatically satisfied since 
${\tilde{K}}_{ab}$ is pure trace and we set ${\tilde{J}}_a=0$. In the de Sitter case we have:
\begin{equation}
a(t)=e^{H_{\Lambda}t},\;\;\;H_{\Lambda}=c\;\sqrt{\frac{\Lambda}{3}},
\label{s1}
\end{equation}
We must satisfy only the Hamiltonian constraint that
for the background metric is trivially given by:
\begin{equation}
H^2=\frac{c^2\Lambda}{3}.
\label{s2}
\end{equation}
By following the same reasoning present in \cite{12}, in the presence of a spherical perturbation, the Hamiltonian constraint becomes
\begin{equation}
R^{(3)}[g]-K_{ab}K^{ab}+Tr(K_{ab})^2=16\pi G\rho,
\end{equation}
where $R^{(3)}$ is the Ricci scalar on $\Sigma$ and:
\begin{equation}
\rho={\rho}_{\Lambda}+\delta{\rho}_S,\;\;
{\rho}_{\Lambda}=\frac{\Lambda c^2}{8\pi G}.
\label{s3}
\end{equation}
Also in this case, under the condition that $K_{a}^{a}=const$, i.e. the rate of expansion of  the volume is not perturbed
($\delta J_b=0$), is supposed to hold. As a result, the Hubble rate of the perturbed metric is left unchanged.
In this way,  thanks to the form of the spatial part of the de Sitter unperturbed metric (${\tilde{g}}_{ab}=a^2(t){\delta}_{ab}$),
the only difference with the calculations present in \cite{12} is that the mass excess is calculated 
with respect to the 'unperturbed' density ${\rho}_{\Lambda}$. 
Hence, instead of (\ref{3}) we can write:
\begin{equation}
\frac{G}{c^2}\delta M_S < \frac{L_S}{2}+\frac{A}{4\pi}\sqrt{\frac{\Lambda}{3}}.
\label{s4}
\end{equation}
By following the reasonings of \cite{9}, we obtain again expression (\ref{1}) with $H$ given by (\ref{s2}).

\section*{Appendix 2}
As stated at the end of section 4, if we consider quantum effects, the introduction
of $\hbar$ can lead to a constant non-zero value for $U_h$ in presence of a cosmological constant. In this case, 
after introducing an effective energy density ${\rho}_h$
at the Hubble radius of our universe, we can write
$U_h=c^2{\rho}_h\frac{4\pi c^3}{3H^3}$: we have
\begin{equation} 
U=const.=K=c^5{\rho}_h\frac{4\pi}{3H^3}.
\label{b17}
\end{equation} 
Equation (\ref{b17}) implies that ${\rho}_h\sim H^3$. More precisely, consider
an initial time $t_i$ with $H(t_i)=H_i, {\rho}_h(t_i)={\rho}_{hi}$: we have
\begin{equation}
{\rho}_h(t)={\rho}_{hi}\frac{H^3}{H_i^3}.
\label{b22}
\end{equation}
Since it is  expected that the semiclassical approximation leading to (\ref{1}) is certainly valid after the Planck era, we expect the time $t_i$ be near the inflationary epoch, where quantum effects on the geometry are negligible. 
Suppose now that the universe is dominated  by this effective
energy-density ${\rho}_h$. By inserting ${\rho}_h$ into Einstein's equation $H^2=8\pi G/3({\rho}_h)$ we obtain 
$H=3/(8\pi G)H_i^3/{\rho}_i$, i.e. a constant value near to the Planck scale. This simple fact suggests that, if the matter 
content of the universe were dominated at some time by
${\rho}_h$, then a de Sitter phase begins.

We now investigate the possible functional relation between ${\rho}_h$ and the usual energy density content of the universe.
Suppose that the universe is filled with usual matter of some species ${\rho}_s(t)$. Friedmann equations
dictate that $H^2=8\pi G/3 {\rho}_s$. From (\ref{b22}) we obtain
\begin{equation} 
{\rho}_h=B_i{\rho}_s(t) H(t),\;\;B_i=\frac{{\rho}_{hi}}{{\rho}_{si} H_i}.
\label{b23}
\end{equation}
Equation (\ref{b23}) shows a coupling between ordinary matter and ${\rho}_h$. This coupling is negligible at late times but is 
huge soon after the Planck era. Suppose that the primordial time $t_i$ is the begin of the primordial inflation $t_I$. 
The expression (\ref{b23}) is similar to the ones present in the so called bulk cosmology \cite{e3} related to the entropic force
(in particular, also in \cite{e1,e2,e3}, an 'entropic pressure'  proportional to $H^2$ in the Friedmann's equations 
is analyzed).
After introducing the density parameters for the species $s$ as ${\Omega}_s=8\pi{\rho}_s/(3H^2)$ and by denoting 
with the subscript ''$0$'' the actual time, we have:
\begin{equation} 
\frac{{\Omega}_{h0}}{{\Omega}_{s0}}=\frac{{\Omega}_{hI}}{{\Omega}_{sI}}\frac{H_0}{H_I}.
\label{b24}
\end{equation}
If we identify the parameter ${\Omega}_{h0}$  with the actual dark energy component and with 
the parameter ${\Omega}_{s0}$ the
actual dark matter component, we have $\frac{{\Omega}_{h0}}{{\Omega}_{s0}}\simeq 2.3$. 
Concerning the ratio $H_0/H_I$, from theoretical estimates (see \cite{21}) we have $H_I/H_0\sim (10^{20},10^{61})$.
By putting these values in (\ref{b24}), we obtain $\frac{{\Omega}_{hI}}{{\Omega}_{sI}}\sim (10^{20},10^{61})$.
As a consequence, if primordial inflation is due to ${\rho}_h$, then the huge ratio 
$\frac{{\Omega}_{hI}}{{\Omega}_{sI}}$ can account for the actual value of $H_0$ and also for
the actual value of the cosmological constant.

\end{document}